\documentclass[sigconf]{acmart}
\usepackage{booktabs} 
\usepackage{enumitem}
\usepackage{makecell}
\usepackage{textcomp}
\usepackage{amsmath}
\usepackage{algorithm}
\usepackage[noend]{algpseudocode}
\usepackage{pdfpages}
\usepackage{graphicx}
\usepackage{subfigure}
\usepackage{amssymb}
\usepackage{mathrsfs}
\usepackage{xcolor}
\usepackage{epsfig}
\usepackage{graphicx}

\copyrightyear{2020}
\acmYear{2020}
\setcopyright{acmcopyright}
\acmConference[WSDM '20] {The Thirteenth ACM International Conference on Web Search and Data Mining}{February 3--7, 2020}{Houston, TX, USA}
\acmBooktitle{The Thirteenth ACM International Conference on Web Search and Data Mining (WSDM'20), February 3--7, 2020, Houston, TX, USA}
\acmPrice{15.00}
\acmDOI{10.1145/3336191.3371769}
\acmISBN{978-1-4503-6822-3/20/02}

\begin{document}
\fancyhead{}

\title{Estimation--Action--Reflection: Towards Deep Interaction Between Conversational and Recommender Systems}








\author{Wenqiang Lei$^1$, Xiangnan He$^2$$^*$, Yisong Miao$^1$, Qingyun Wu$^3$, Richang Hong$^4$, Min-Yen Kan$^1$, Tat-Seng Chua$^1$}

\thanks{$^*$Xiangnan He is the Corresponding Author}

\affiliation{\institution{$^1$National University of Singapore, $^2$University of Science and Technology of China\\
$^3$University of Virginia, $^4$Hefei University of Technology}}

\email{wenqianglei@gmail.com,  xiangnanhe@gmail.com, miaoyisong@gmail.com, qw2ky@virginia.edu}
\email{hongrc@hfut.edu.cn, kanmy@comp.nus.edu.sg, chuats@comp.nus.edu.sg}

\settopmatter{printacmref=false, printfolios=false}



\begin{abstract}
Recommender systems are embracing conversational technologies to obtain user preferences dynamically, and to overcome inherent limitations of their static models.
A successful \textit{Conversational Recommender System} (CRS) requires proper handling of interactions between conversation and recommendation.  We argue that three fundamental problems need to be solved: 1) what questions to ask regarding item attributes, 2) when to recommend items, and 3) how to adapt to the users' online feedback.  To the best of our knowledge, there lacks a unified framework that addresses these problems.

In this work, we fill this missing interaction framework gap by proposing a new CRS framework named \textit{Estimation--Action--Reflection}, or {EAR}, which consists of three stages to better converse with users. (1) Estimation, which builds predictive models to estimate user preference on both items and item attributes; (2) Action, which learns a dialogue policy to determine whether to ask attributes or recommend items, based on Estimation stage and conversation history; and (3) Reflection, which updates the recommender model when a user rejects the recommendations made by the Action stage.
We present two conversation scenarios on binary and enumerated questions, and conduct extensive experiments on two datasets from Yelp and LastFM, for each scenario, respectively.
Our experiments demonstrate significant improvements over the state-of-the-art method CRM~\cite{Sun:2018:CRS:3209978.3210002}, corresponding to fewer conversation turns and a higher level of recommendation hits. 
\end{abstract}


\maketitle

\begingroup
    \fontsize{8pt}{8pt}\selectfont
        \textbf{ACM Reference Format:}\\
        Wenqiang Lei, Xiangnan He, Yisong Miao, Qingyun Wu, Richang Hong, Min-Yen Kan, \& Tat-Seng Chua. 2020. Estimation--Action--Reflection: Towards Deep Interaction Between Conversational and Recommender Systems. In \textit{The Thirteenth ACM International Conference on Web Search and Data Mining (WSDM'20), February 3--7, 2020, Houston, TX, USA.} ACM, NY, NY, USA, 9 pages. https://doi.org/10.1145/3336191.3371769
\endgroup

\section{Introduction}
\label{sec:introduction}


Recommender systems are emerging as an important means of facilitating users' information seeking~\cite{MF,BPR,NCF,ACF}. However, much of such prior work in the area solely leverages the offline historical data to build the recommender model (henceforth, the {\it static recommender system}). This offline focus causes the recommender to suffer from an inherent limitation in the optimization of offline performance, which may not necessarily match online user behavior. User preference can be diverse and often drift with time; and as such, it is difficult to know the exact intent of a user when he uses a service even when the training data is sufficient. 


The rapid development of conversational techniques~\cite{nips18/DeepConv,Liao:2018,sigir18/chatmore,acl18/sequicity,jin2018explicit} brings an unprecedented opportunity that allows a recommender system to dynamically obtain user preferences through conversations with users. This possibility is envisioned as the {\it conversational recommender system} (CRS), for which the community has started to expend effort in exploring its various settings. \cite{zhang2018towards} built a conversational search engine by focusing on document representation. 
\cite{nips18/DeepConv} developed a dialogue system to suggest movies for cold start users, contributing to language understanding and generation for the purpose of recommendation, but does not consider modeling users' interaction histories (e.g., clicks, ratings).
In contrast, \cite{christakopoulou2018q} does considers user click history in recommending, but their CRS only handles single-round recommendation. That is, their model considers a scenario in which the CRS session terminates after making a single recommendation, regardless of whether the recommendation is satisfactory or not. While a significant advance, we feel this scenario is unrealistic in actual deployments.

In particular, we believe CRS models should inherently adopt a {\it multi-round} setting: a CRS converses with a user to recommend items based on his click history (if any). At each round, the CRS is allowed to choose two types of actions --- either explicitly asking whether a user likes a certain item attribute or recommending a list of items. In a session, the CRS may alternate between these actions multiple times, with the goal of finding desirable items while minimizing the number of interactions.
This multi-round setting is more challenging than the single-round setting, as the CRS needs to strategically plan its actions. The key in performing such planning, from our perspective, lies in the interaction between the conversational component (CC; responsible for interacting with the user) and the recommender component (RC; responsible for estimating user preference -- e.g., generating the recommendation list). We summarize three fundamental problems toward the deep interaction between CC and RC as follows:
\begin{itemize}[leftmargin=1mm]
    \item \textit{What attributes to ask?} 
    A CRS needs to choose which attribute to ask the user about. For example, in music recommendation, it may ask ``Would you like to listen to classical music?'', expecting a binary yes/no response\footnote{Note that it is possible to compose questions eliciting an enumerated response; i.e., ``Which music genre would you consider? I have pop, funk ...''. However, this is a design choice depending on the domain requirements. In describing our method, we consider the basic single-attribute case.  However in experiments, we also justify the effectiveness of EAR in asking such enumerated questions on Yelp. For the purpose of exposition, we have chosen to avoid open questions that do not constrain user response for now.  Even interpreting user responses to such questions is considered a challenging task~\cite{chen2018hierarchical}.}. 
    If the answer is ``yes'', it can focus on items containing the attribute, benefiting the RC by reducing uncertainty in item ranking. However, if the answer is ``no'', the CRS expends a conversation turn with less gain to the RC. To achieve the goal of hitting the right items in fewer turns, the CC must consider whether the user will like the asked attribute. This is exactly the job of the RC which scrutinizes the user's historical behavior. 

    \item \textit{When to recommend items?} 
    With sufficient certainty, the CC should push the recommendations generated by the RC. 
    A good timing to push recommendations should be when 1) the candidate space is small enough; when 2) asking additional questions is determined to be less useful or helpful, from the perspective of either information gain or user patience; and when 3) the RC is confident that the top recommendations will be accepted by the user. Determining the appropriate  timing should take both the conversation history of the CC and the preference estimation of the RC into account. 
    \item \textit{How to adapt to users' online feedback?} After each turn, the user gives feedback; i.e., ``yes''/``no'' to a queried attribute, or an ``accept''/``reject'' the recommended items. (1) For ``yes'' on the attribute, both user profile and item candidates need to be updated to generate better recommendations; this requires the offline RC training to take such updates into account. (2) For ``no', the CC needs to adjust its strategy accordingly. (3) If the recommended items are rejected, the RC model needs to be updated to incorporate such a negative signal. Although adjustments seem only to impact either the RC or the CC, we show that such actions impact both.  
\end{itemize}

Towards the deep interaction between CC and RC, we propose a new solution named \textit{Estimation--Action--Reflection} (EAR), which consists of three stages. Note that the stages do not necessarily align with each of the above problems. 
(a) Estimation, which builds predictive models offline to estimate user preference on items and item attributes. 
Specifically, we train a factorization machine~\cite{rendle2010factorization}~(FM) using user profiles and item attributes as input features. Our Estimation stage builds in two novel advances: 1) the joint optimization of FM on the two tasks of item prediction and attribute prediction, and 2) the adaptive training of conversation data with online user feedback on attributes. 
(b) Action, which learns 
the conversational strategy that determines whether to ask or recommend, and what attribute to ask. We train a policy network with reinforcement learning, optimizing the reward of shorter turns and successful recommendations based on the FM's estimation of user preferred items and attributes, and the dialogue history. 
(c) Reflection, which adapts the CRS with user's online feedback. Specifically, when a user rejects the recommended items, we construct new training triplets by treating the items as negative instances and update the FM in an online manner.  
In summary, the main contributions of this work are as follows:
\begin{itemize}[leftmargin=*]
    \item We comprehensively consider a multi-round CRS scenario that is more realistic than previous work, highlighting the importance of researching into the interactions between the RC and CC to build an effective CRS. 
    \item We propose a three-stage solution, EAR, integrating and revising several RC and CC techniques to construct a solution that works well for the conversational recommendation. 
    \item We build two CRS datasets by simulating user conversations to make the task suitable for offline academic research. We show our method outperforms several state-of-the-art CRS methods and provide insight on the task. 
\end{itemize}

\vspace{-1mm}
\section{Multi-round Conversational Recommendation Scenario}\label{sec:pre}


\begin{figure}[t]
\centering
\includegraphics[height=5cm]{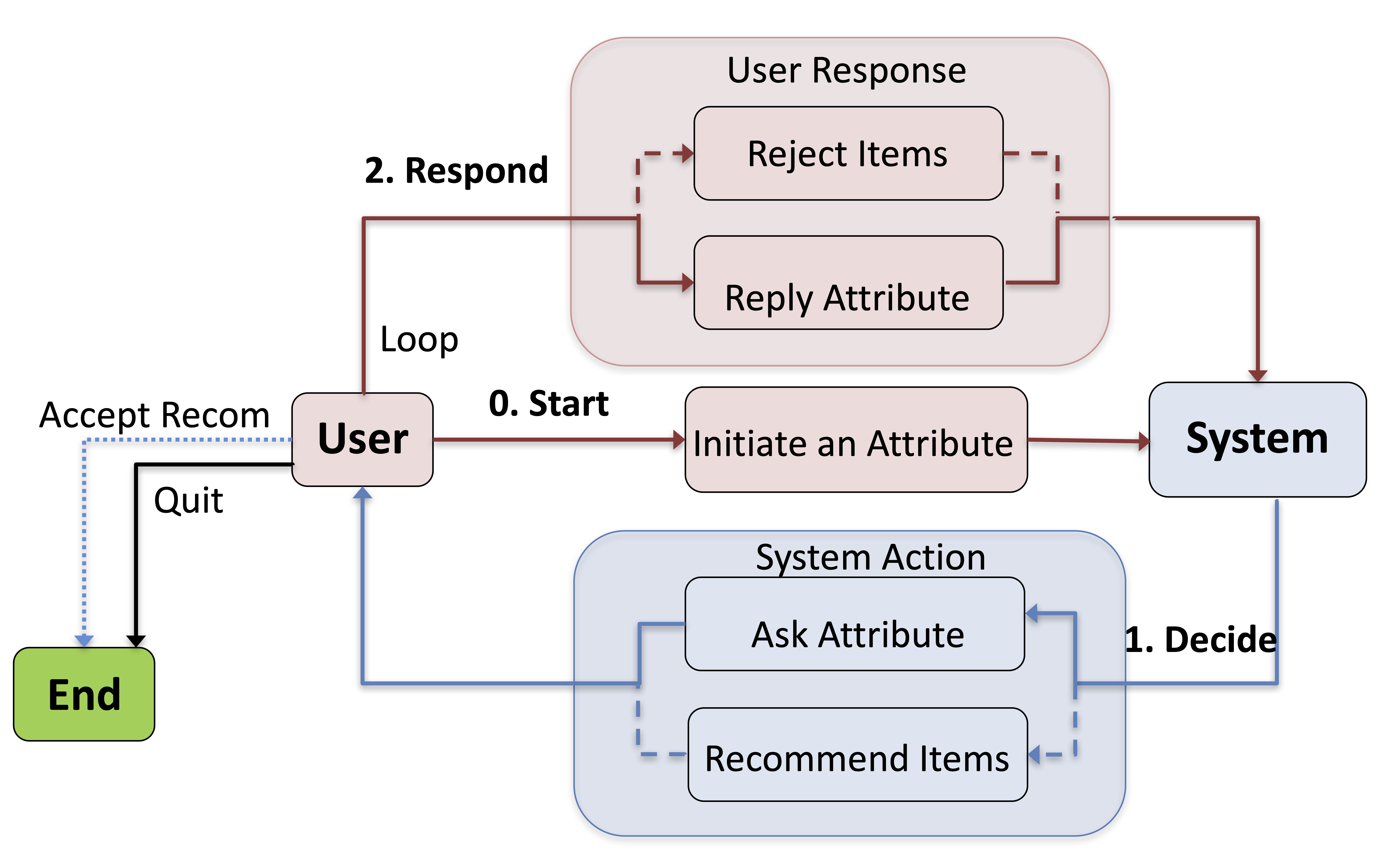}\vspace{-10pt}
\caption{The workflow of our multi-round conversational recommendation scenario. The system may recommend items multiple times, and the conversation ends only if the user accepts the recommendation or chooses to quit.}\vspace{-10pt}
\label{fig:preli_workfolow}
\vspace{-10pt}
\end{figure}

Following~\cite{christakopoulou2018q}, we denote one trial of recommendation as a \emph{round}. This paper considers conversational recommendation as an inherently \textit{multi-round} scenario, where a CRS interacts with the user by asking attributes and recommending items multiple times until the task succeeds or the user leaves. To distinguish the two, we term the setting \emph{single-round} where the CRS only makes recommendations once, ending the session regardless of the outcome, as in \cite{Sun:2018:CRS:3209978.3210002,christakopoulou2018q}. 

We now introduce the notation used to formalize our setting.
Let $u\in\mathcal{U}$ denote a user $u$ from the user set $\mathcal{U}$ and $v\in\mathcal{V}$ denote an item $v$ from the item set $\mathcal{V}$. Each item $v$ is associated with a set of attributes $\mathcal{P}_v$ which describe its properties, such as music genre ``classical'' or ``jazz'' for songs in LastFM, or tags such as ``nightlife'', ``serving burgers'', or ``serving wines'' for businesses in Yelp. We denote the set of all attributes as $\mathcal{P}$ and use $p$ to denote a specific attribute. 
Following~\cite{Sun:2018:CRS:3209978.3210002,zhang2018towards}, a CRS session is started with $u$'s specification of a preferred attribute $p^0$,
then the CRS filters out candidate items that contain the preferred attribute $p^0$.
Then in each turn $t$ ($t=1, 2, ..., T$; $T$ denotes the last turn of the session), the CRS needs to choose an action: \textit{recommend} or \textit{ask}:
\begin{itemize}[leftmargin=*]
    \item If the \textsc{action} is \emph{recommend}, we denote the recommended item list $\mathcal{V}^t\subset\mathcal{V}$ and the action as $a_{rec}$. Then the user examines whether $\mathcal{V}^t$ contains his desired item. If the feedback is positive, this session succeeds and can be terminated. Otherwise, we mark $\mathcal{V}^t$ as \textit{rejected} and move to the next round.
    \item If the \textsc{action} is \emph{ask} (where the asked attribute is denoted as $p^t\in\mathcal{P}$ and the action as $a_{ask}(p^t)$), the user states whether he prefers items that contain the attribute $p^t$ or not. If the feedback is positive, we add $p^t$ into $\mathcal{P}_u$ to denote the preferred attributes the user in the current session. Otherwise, we mark $p^t$ as \textit{rejected}; regardless of rejection or not, we move to the next turn. 
\end{itemize}




\noindent 
This whole process naturally forms a interaction loop (Figure~\ref{fig:preli_workfolow}) where the CRS may ask zero to many questions before making recommendations. 
A session terminates if a user accepts the recommendations or leaves due to his impatience.  We set the main goal of the CRS as making desired recommendations within as few rounds as possible.

\section{Proposed Methods}
EAR consists of a recommendation and conversation component (RC and CC) which interact intensively in the three--stage conversational process.
The system starts working at the \emph{estimation} stage where the RC ranks candidate items and item attributes for the user, so as to support the action decision of the CC.
After the \emph{estimation} stage, the system moves to the \emph{action} stage where the CC decides whether to choose an attribute to ask, or make a recommendation according to the ranked candidates and attributes, and the dialogue history.
If the user likes the attribute asked by the RC, the CC feeds this attribute back to the RC to make a new \emph{estimation} again; otherwise, the system stays at the \emph{action} stage: updates the dialogue history and chooses another action. Once a recommendation is rejected by a user, the CC sends the rejected items back to RC, triggering the \emph{reflection} stage where the RC adjusts its estimations.
After that, the system enters the \emph{estimation} stage again.


\subsection{Estimation}
As discussed before, the multi-round conversational scenario brings in new challenges to the traditional RC.
Specifically, the CC interacts with a user $u$ and accumulates evidence on
his
preferred attributes, denoted as  $\mathcal{P}_u=\{p_1,p_2,..,p_n\}$\footnote{We detail how to obtain such data in experiments Section~\ref{sec:simu}.}. 
Importantly, different from traditional recommendation methods~\cite{BPR,NCF}, the RC here needs to make full use of $\mathcal{P}_u$ aiming to accurately predict both user's the preferred items and preferred attributes. 
These two goals exert positive influence on EAR, where the first directly contributes to success rate of recommendation, and the second guides the CC to choose better attributes to ask users so as to shorten the conversation.
In the following, we first introduce the basic form of the recommendation method, followed by detail on how we adapt our proposed method to achieve both goals simultaneously.

\subsubsection{Basic Recommendation Method}
we choose the factorization machine (FM)~\cite{rendle2010factorization} as our predictive model due to its success and wide usage in recommendation tasks. However, FM considers all pairwise interactions between input features, which is costly and may introduce undesired interactions that negatively affect our two goals. Thus, we only keep the interactions that are useful to our task and remove the others. Given user $u$, his preferred attributes in the conversation $\mathcal{P}_u$, and the target item $v$, we predict how likely $u$ will like $v$ in the conversation session as:
\begin{equation}
\begin{aligned}
    \hat{y}(u,v,\mathcal{P}_u)= \mathbf{u}^T\mathbf{v}
    + \sum_{p_i\in\mathcal{P}_u} \mathbf{v}^T \mathbf{p_i},
\end{aligned}
\end{equation}
\noindent where $\textbf{u}$ and $\textbf{v}$ denote the embedding for user $u$ and item $v$, respectively, and $\textbf{p}_i$ denotes the embedding for attribute $p_i\in \mathcal{P}_u$. Bias terms are omitted for clarity.
The first term $\mathbf{u}^T\mathbf{v}$ models the general interest of the user on the target item, a common term in FM model~\cite{NCF}.
The second term $\sum \mathbf{v}^T \mathbf{p_i}$ models the affinity between the target item and user preferred attributes.
We have also tried to include $v$'s attributes $\mathcal{P}_v$ into FM, but found it brings no benefits. One possible reason is that the item embedding $\textbf{v}$ may have already encoded its attribute information. Thus we also omit it.

To train the FM, we optimize the pairwise Bayesian Personalized Ranking (BPR)~\cite{BPR} objective.
Specifically, given a user $u$, it assumes the interacted items (e.g., visited restaurants, listened music) should be assigned higher scores than those not interacted with.
The loss function of traditional BPR is:
\begin{equation}
    L_{bpr} = \sum_{(u,v,v')\in \mathcal{D}_1} -\mbox{ln} \sigma(\hat{y}(u,v,\mathcal{P}_u) - \hat{y}(u,v',\mathcal{P}_u)) + \lambda_{\Theta} \left\|\Theta\right\|^2
\end{equation}
where $\mathcal{D}_1$ is the set of pairwise instances for BPR training, $\mathcal{D}_1 := \{(u, v, v') \mid v' \in \mathcal{V}_{u}^- \}$, where $v$ is the interacted item of the conversation session (i.e., the ground truth item of the session), $\mathcal{V}_{u}^{-}$ := $\mathcal{V} \backslash \mathcal{V}_{u}^{+}$ denotes the set of non-interacted items of user $u$ and $\mathcal{V}_{u}^{+}$ denotes the items interacted by $u$.
$\sigma$ is the sigmoid function, and $\lambda_{\Theta}$ is the regularization parameter to prevent overfitting.

\subsubsection{Attribute-aware BPR for Item Prediction.} However, in our scenario, the emphasis of CRS is to rank the items that contain the user preferred attributes well.
For example, if $u$ specifies ``Mexican restaurant'' as his preferred attribute, a good CRS needs to rank his preferred restaurants among all available Mexican restaurants. 
To capture this, we propose to sample two types of negative examples:
\begin{equation}
\begin{aligned}
    \mathcal{V}_{u}^{-} := \mathcal{V} \backslash \mathcal{V}_{u}^{+}, \quad
    \widehat{\mathcal{V}}_{u}^- := \mathcal{V}_{cand} \backslash \mathcal{V}_{u}^{+},
\end{aligned}
\end{equation}
where $\mathcal{V}_{u}^{-}$ is the same negative samples as in the traditional BPR setting, i.e., all non-interacted items of $u$.
$\mathcal{V}_{cand}$ denotes the current candidate items satisfying the partially known preference $\mathcal{P}_u$ in the conversation, and $\widehat{\mathcal{V}}_{u}^-$ is the subset of $\mathcal{V}_{cand}$ that excludes the observed items $\mathcal{V}_{u}^{+}$. The two types of pairwise training instances is defined as:
\begin{equation}
    \mathcal{D}_1 := \{(u, v, v') \mid v' \in \mathcal{V}_{u}^- \}, \quad \mathcal{D}_2 := \{(u, v, v') \mid v' \in \widehat{\mathcal{V}}_{u}^- \},
\end{equation}

We then train the FM model by optimizing
both $\mathcal{D}_1$ and $\mathcal{D}_2$:
\begin{equation}
\begin{aligned}
    L_{item} &= \sum_{(u,v,v')\in \mathcal{D}_1} -\mbox{ln} \sigma(\hat{y}(u,v,\mathcal{P}_u) - \hat{y}(u,v',\mathcal{P}_u)) \\
    &+ \sum_{(u,v,v')\in \mathcal{D}_2} -\mbox{ln} \sigma(\hat{y}(u,v,\mathcal{P}_u) - \hat{y}(u,v',\mathcal{P}_u)) + \lambda_{\Theta} \left\|\Theta\right\|^2,
\end{aligned}
\end{equation}
where the first loss learns $u$'s general preference, and the second loss learns $u$'s specific preference given the current candidates. It is worth noting adding the second loss for training is critical for the model ranking well on the current candidates. This is very important for CRS since the candidate items dynamically change with user feedback along the conversation. However, the state-of-the-art method CRM~\cite{Sun:2018:CRS:3209978.3210002} does not account for this factor, being insufficient in considering the interaction between the CC and RC.

\subsubsection{Attribute Preference Prediction.} We formulate the task of the second goal of accurate attribute prediction separately. This prediction of attribute preference is mainly used in the CC to support the action on which attribute to ask ({\it c.f.} Sec~\ref{ss:action}). As such, we take $u$'s preferred attributes in the current session into account:
\begin{equation}\label{eq:sim}
    \hat{g}(p|u,\mathcal{P}_u) = \textbf{u}^T \textbf{p} + \sum_{p_i\in \mathcal{P}_u} \textbf{p}^T \textbf{p}_i,
\end{equation}
which estimates $u$'s preference on attribute $p$, given $u$'s current preferred attributes $\mathcal{P}_u$.
To train the model, we also employ BPR loss, and assume that the attributes of the ground truth item $v$ (of the session) should be ranked higher than other attributes:
\begin{equation}
\begin{aligned}
    L_{attr} = \sum_{(u,p,p')\in \mathcal{D}_3} -\mbox{ln} \sigma(\hat{g}(p|u, \mathcal{P}_u) - \hat{g}(p'|u, \mathcal{P}_u)) + \lambda_{\Theta} \left\|\Theta\right\|^2,
\end{aligned}
\end{equation}
where the pairwise training data $D_3$ is defined as:
\begin{equation}
    D_3 = \{(u, p, p')| p\in \mathcal{P}_v, p'\in \mathcal{P} \backslash \mathcal{P}_v \},
\end{equation}
where $\mathcal{P}_v$ denotes item $v$'s attributes.

\subsubsection{Multi-task Training.}\label{ss:multi-task}
We perform joint training on the two tasks of item prediction and attribute prediction, which has the potential of mutual benefits since their parameters are shared. The multi-task training objective is:
\begin{equation}
    L = L_{item} + L_{attr}.
\end{equation}

\noindent Specifically, we first train the model with $L_{item}$. After it converges, we continue optimizing the model using $L_{attr}$. We iterate the two steps until convergence under both losses.  Empirically, 2-3 iterations are sufficient for convergence.

\subsection{Action}\label{ss:action}


After the estimation stage, the action stage finds the best strategy for \textit{when to recommend}. We adopt reinforcement learning (RL) to tackle this multi-round decision making problem, aiming to accomplish successful recommendation in shorter number of turns.
It is worth noting that since our focus is on conversational recommendation strategy, as opposed to fluent dialogue (the language part), we use templates as wrappers to handle user utterances and system response generation.
That is to say, this work serves as an upper bound study of real applications as we do not include the errors for language understanding and generation. 

\subsubsection{State Vector.} The state vector is a bridge for the interaction between the CC and RC. We encode information from the RC and dialogue history into a state vector, providing it to the CC to choose actions.
The state vector is a concatenation of four component vectors that encode signal from different perspectives:

\vspace{-10pt}
\begin{equation}\label{eq:state}
    \textbf{s} = \textbf{s}_{ent} \oplus \textbf{s}_{pre} \oplus \textbf{s}_{his} \oplus \textbf{s}_{len}.
\end{equation}

Each of the vector components captures an assumption on asking which attribute could be most useful, or whether now is a good time to push a recommendation. They are defined as follows:
\begin{itemize}[leftmargin=1mm]
    \item $\textbf{s}_{ent}$: This vector encodes the entropy information of each attribute among the attributes of the current candidate items $\mathcal{V}_{cand}$. The intuition is that asking attributes with large entropy helps to reduce the candidate space, thus benefits finding desired items in fewer turns. Its size is the attribute space size $|\mathcal{P}|$, where the $i$-th dimension denotes the entropy of the attribute $p_i$.
    \item $\textbf{s}_{pre}$: This vector encodes $u$'s preference on each attribute. It is also of size $|\mathcal{P}|$, where each dimension is evaluated by Equation~(\ref{eq:sim}) on the corresponding attribute. The intuition is that the attribute with high predicted preference is likely to receive positive feedback, which also helps to reduce the candidate space.
    \item $\textbf{s}_{his}$: This vector encodes the conversation history. Its size is the number of maximum turns $T$, where each dimension $t$ encodes user feedback at turn $t$. Specifically, we use -1 to represent recommendation failure, 0 to represent asking an attribute that $u$ disprefers, and 1 to represent successfully asking about an attribute that $u$ desires. This state is useful to determine when to recommend items. For example, if the system has asked about a number of attributes for which $u$ approves, it may be a good time to recommend.
    \item $\textbf{s}_{len}$: This vector encodes the length of the current candidate list. The intuition is that if the candidate list is short enough, EAR should turn to recommending to avoid wasting more turns. We divide the length $|\mathcal{V}_{cand}|$ into ten categorical (binary) features to facilitate the RL training.   
\end{itemize}

\noindent It is worth noting that besides $\textbf{s}_{his}$, the other three vectors are all derived from the RC component.
We claim that this is a key difference from existing conversational systems~\cite{zhang2018towards,nips18/DeepConv,Sun:2018:CRS:3209978.3210002,christakopoulou2018q,Liao:2018}; i.e., the CC needs to take information from the RC to decide the dialogue action. In contrast to EAR, the recent conversational recommendation method CRM~\cite{Sun:2018:CRS:3209978.3210002} makes decisions based only on the belief tracker that records the preferred attributes of the user, which makes it less informative. As such, CRM is less effective especially when the number of attributes is large (their experiments only deal with 5 attributes, which is insufficient for real-world applications).

\subsubsection{Policy Network and Rewards} The conversation action is chosen by a policy network in our CC. In order to demonstrate the efficacy of our designed state vector, we purposely choose a simple policy network --- a two-layer multi-layer perceptron, which can be optimized with the standard policy gradient method.
It contains two fully-connected layers
and maps the state vector $\textbf{s}$ into the action space. The output layer is normalized to be a probability distribution over all actions by $softmax$.
In terms of the action space, we follow the previous method \cite{Sun:2018:CRS:3209978.3210002}, which includes all attributes $\mathcal{P}$ and a dedicated action for recommendation. To be specific, we define the action space as $\mathcal{A} = \{ a_{rec} \cup \{ a_{ask}(p) | p\in \mathcal{P}\} \},$
which is of size $|\mathcal{P}|+1$.
After the CC takes an action at each turn, it will receive an immediate reward from the user (or user simulator).
This will guide the CC to learn the optimal policy that optimizes long-term reward. In EAR, we design four kinds of rewards, namely: (1) $r_{suc}$, a strongly positive reward when the recommendation is successful, (2) $r_{ask}$, a positive reward when the user gives positive feedback on the asked attribute, (3) $r_{quit}$, a strongly negative reward if the user quits the conversation, (4) $r_{prev}$, a slightly negative reward on every turn to discourage overly lengthy conversations.
The intermediate reward $r_t$ at turn $t$ is the sum of the above four rewards, $r_t=r_{suc}+r_{ask}+r_{quit}+r_{prev}$.

We denote the policy network as $\pi(a^t\mid\textbf{s}^t)$, which returns the probability of taking action $a^t$ given the state $\textbf{s}^t$. Here $a^t\in \mathcal{A}$ and $\textbf{s}^t$ denote the action to take and the state vector of the $t$-th turn, respectively.
To optimize the policy network, we use the standard policy gradient method~\cite{REINFORCE}, formulated as follows:
 \begin{equation}
    \theta \leftarrow \theta - \alpha \bigtriangledown \mbox{log} \pi_{\theta}(a^t\mid \textbf{s}^t) R_{t},
\end{equation}
where $\theta$ denotes the parameter of the policy network, $\alpha$ denotes the learning rate of the policy network, and $R_{t}$ is the total reward accumulating from turn $t$ to the final turn $T$: $R_t = \sum_{t'=t}^{T} \gamma^{T-t'}r_{t'},$
where $\gamma$ is a discount factor which discounts future rewards over immediate reward.

\subsection{Reflection}
This stage also implements the interaction between the CC and RC. It is triggered when the CC pushes the recommended items $\mathcal{V}^{t}$ to the user but gets rejected, so as to update the RC model for better recommendations in future turns.
In the traditional static recommender system training scenario~\cite{BPR,NCF}, one issue is the absence of true negative samples, since users do not explicitly indicate what they dislike.
In our conversational case, the rejection feedback is an explicit signal on user dislikes which are highly valuable to utilize; moreover, it indicates that the offline learned FM model improperly assigns high scores to the rejected items.
To leverage on this source of feedback, we treat the rejected items $\mathcal{V}^{t}$ as negative samples, constructing more training examples to refresh the FM model. Following the offline training process, we also optimize the BPR loss:
\begin{equation}
    L_{ref} = \sum_{(u,v,v')\in \mathcal{D}_4} -\mbox{ln} \sigma(\hat{y}(u,v,\mathcal{P}_u) - \hat{y}(u,v',\mathcal{P}_u)) + \lambda_{\Theta} \left\|\Theta\right\|^2
\end{equation}
where $\mathcal{D}_4 := \{(u, v, v') \mid v \in \mathcal{V}_{u}^{+} \wedge v' \in \mathcal{V}^{t}\}$. Note that this stage is performed in an online fashion, where we do not have access to the ground truth positive item. Thus, we treat the historically interacted items $\mathcal{V}^+_u$ as the positive items to pair with the rejected items. We put all examples in $D_4$ into a batch and perform batch gradient descent. Empirically, it takes 3-5 epochs to converge, sufficiently efficient for online use.

Note that although it sounds reasonable to also update the policy network of the CC (since the rejection feedback implies that it is not an appropriate timing to push recommendation), we currently do not perform this due to high difficulty of online updating RL agent and leave it for future work. 


\section{Experiments}
EAR \footnote{Datasets, source code and demos at our project homepage: https://ear-conv-rec.github.io} is built based on the guiding ideology of interaction between the CC and RC. To validate this ideology, we first evaluate the whole system to examine the overall effect brought by the interaction. Then, we perform ablation study to investigate the effect of interaction on each individual component. Specifically, we have the following research questions (RQ) to guide experiments on two datasets.
\begin{itemize}[leftmargin=*]
	\item \textbf{RQ1.} How is the overall performance of EAR comparing with existing conversational recommendation methods?
	\item \textbf{RQ2.} How do the attribute-aware BPR and multi-task training of the \textit{estimation} stage contribute to the RC?
	\item \textbf{RQ3.} Is the state vector designed for the CC in the \textit{action} stage appropriate?	
	\item \textbf{RQ4.} Is the online model update of the \textit{reflection} stage useful in obtaining better recommendation?
\end{itemize}

\subsection{Settings}
\subsubsection{Datasets}
We conduct experiments on two datasets: Yelp\footnote{https://www.yelp.com/dataset/} for business recommendation and LastFM\footnote{https://grouplens.org/datasets/hetrec-2011/} for music artist recommendation.
First, we follow the common setting of recommendation evaluation~\cite{NCF,BPR} that reduces the data sparsity by pruning the users that have less than 10 reviews. We split the user--item interactions in the ratio of 7:2:1 for training, validation and testing.
Table~\ref{tab:data} summarizes the statistics of the datasets.
\begin{table}[t]
\caption{Dataset statistics.}
\vspace{-5pt}
\begin{tabular}{|l|l|l|l|l|}
\hline\small
\textbf{Dataset} & \#\textbf{users} & \#\textbf{items} & \#\textbf{interactions} & \#\textbf{attributes}  \\ \hline
Yelp & 27,675      & 70,311  & 1,368,606 & 590     \\ \hline
LastFM            & 1,801      &  7,432 & 76,693 & 33     \\ \hline
\end{tabular}
\label{tab:data}
\vspace{-5pt}
\end{table}

For the item attributes, we preprocess the original attributes of the datasets by merging synonyms and eliminating low frequency attributes, resulting in 590 attributes in Yelp and 33 attributes in LastFM. In real applications, asking about attributes in a large attribute space (e.g., on Yelp dataset) causes overly lengthy conversation. We therefore consider both the binary question setting (on LastFM) and enumerated question (on Yelp). To enable the enumerated question setting, we build a two-level taxonomy on the attributes of the Yelp data. For example, the \textit{parent attribute} of  \{``wine", ``beer", ``whiskey''\} is ``alcohol''. We create 29 such parent attributes on the top of the 590 attributes, such as ``nightlife'', ``event planning \& services'', ``dessert types'' etc. In the enumerated question setting, the system choose one parent attribute to ask. This is to say, we change the size of the output space of the policy network to be $29+1=30$. At the same time, it also displays all its child attributes and ask the user to choose from them (the user can reply with multiple child attributes).
Note that choosing what kinds of questions to ask is an engineering design choice by participants, here we evaluate our model on both settings. 

\subsubsection{User Simulator For Multi-round Scenario.}
\label{sec:simu}
Because the conversational recommendation is a dynamic process, we follow  ~\cite{zhang2018towards,Sun:2018:CRS:3209978.3210002}) to create a user simulator to enable the CRS training and evaluation.
We simulate a conversation session for each observed interaction between users and items. Specifically, given an observed user--item interaction $(u,v)$, we treat the $v$ as the ground truth item to seek for and its attributes $\mathcal{P}_v$ as the oracle set of attributes preferred by the user in this session. At the beginning, we randomly choose an attribute from the oracle set as the user's initialization to the session. Then the session goes in the loop of the ``model acts -- simulator response" process as introduced in Section~\ref{sec:pre}.
We set the max turn $T$ of a session to 15 and standardize the recommendation list length $\mathcal{V}^{t}$ as 10.

\subsubsection{Training Details}
Following CRM~\cite{Sun:2018:CRS:3209978.3210002},
the training process is divided into offline and online stages. The offline training is to build the RC (i.e., FM) and initialize the policy network (PN) by letting them optimize performance with the offline dialogue history. Due to the scarcity of the conversational recommendation dialogue history, we follow CRM~\cite{Sun:2018:CRS:3209978.3210002} to simulate dialogue history by building a rule-based CRS to interact with the simulator introduced in Section~\ref{sec:simu}. Specifically, the strategy for determining which attribute to ask about is to choose the
attribute
with the maximum entropy. Each turn, the system chooses the recommendation action with probability $10 / max({|\mathcal{V}|, 10})$ where $\mathcal{V}$ is the current candidate set. The intuition is that the confidence of recommendation grows when the candidate size is smaller. We train the RC to give the ground-truth item and oracle attributes higher ranks given the attribute confirmed by users in dialogue histories, while training the policy to mimic the rule-based strategy on the history. Afterwards, we conduct online training, optimizing the PN by letting EAR interact with the user simulator through reinforcement learning.

We tuned all hyper-parameters on the validation set, and empirically set them as followed:
The embedding size of FM is set as 64.
We employ the multi-task training mechanism to optimize FM as described in Section~\ref{ss:multi-task}, using SGD with a regularization strength of 0.001. The learning rate for the first task (item prediction) and second task (attribute prediction) is set to 0.01 and 0.001, respectively.
The size of the two hidden layers in the PN is set as 64.
When the pre-trained model is initialized, we use the REINFORCE algorithm to train the PN.
The four rewards are set as: $r_{suc}$=1, $r_{ask}$=0.1, $r_{quit}$=-0.3, and $r_{prev}$=-0.1, and the learning rate $\alpha$ is set as $0.001$. The discount factor $\gamma$ is set to be 0.7.

\subsubsection{Baselines.}
As our multi-round conversational recommendation scenario is new, there are few suitable baselines.  We compare our overall performance with the following three:
\begin{itemize}[leftmargin=1mm]
    \item \textbf{Max Entropy}. This method follows the rule we used to generate the conversation history in Section~\ref{sec:simu}. Each turn it asks the attribute that has the maximum entropy among the candidate items. It is claimed in~\cite{dhingra2017towards} that maximum entropy is the best strategy when language understanding is precise. It's worth noting that, in enumerated question setting, the entropy of an attribute is calculated as the sum of its child attributes in the taxonomy (similar approach for attribute preference calculation).
    \item \textbf{Abs Greedy}~\cite{christakopoulou2016towards}. This method recommends items in every turn without asking any question. Once the recommendation is rejected, it updates the model by treating the rejected items as negative examples. According to~\cite{christakopoulou2016towards}, this method achieves equivalent or better performance than popular bandit algorithms like Upper Confidence Bounds~\cite{auer2002using} and Thompson Sampling~\cite{chapelle2011empirical}.
    \item \textbf{CRM}~\cite{Sun:2018:CRS:3209978.3210002}. This is a state-of-the-art CRS. Similar to EAR, it integrates a CC and RC by feeding the belief tracker results to FM for item prediction, without considering much interactions between them. It is originally designed for single-round recommendation. To achieve a fair comparison, we adapt it to the multi-round setting by following the same offline and online training of EAR.
\end{itemize}
It is worth noting that although there are other recent conversational recommendation methods~\cite{zhang2018towards,nips18/DeepConv,christakopoulou2016towards,Liao:2018}, they are ill-suited for comparison due to their different task settings. For example, \cite{zhang2018towards} focuses on document representation which is unnecessary in our case. It also lacks the conversation policy component to decide when to make what action.  \cite{nips18/DeepConv} focuses more on language understanding and generation. We summarize the settings of these methods in Table~\ref{tab:CRS} and discuss differences in Section~\ref{sec:related}.
\subsubsection{Evaluation Metrics}
We use the success rate (SR@t)~\cite{Sun:2018:CRS:3209978.3210002} to measure the ratio of successful conversations, i.e., recommend the ground truth item by turn $t$. 
We also report the average turns (AT) needed to end the session. Larger SR denotes better recommendation and smaller AT denotes more efficient conversation. When studying RC model of offline training, we use the AUC score which is a surrogate of the BPR objective~\cite{BPR}. We conduct one-sample paired t-test to judge statistical significance.

\subsection{Performance Comparison~(RQ1)}

\vspace{-15pt}

\begin{figure}[htbp]
\begin{minipage}[t]{0.47\linewidth}
    \includegraphics[width=\linewidth]{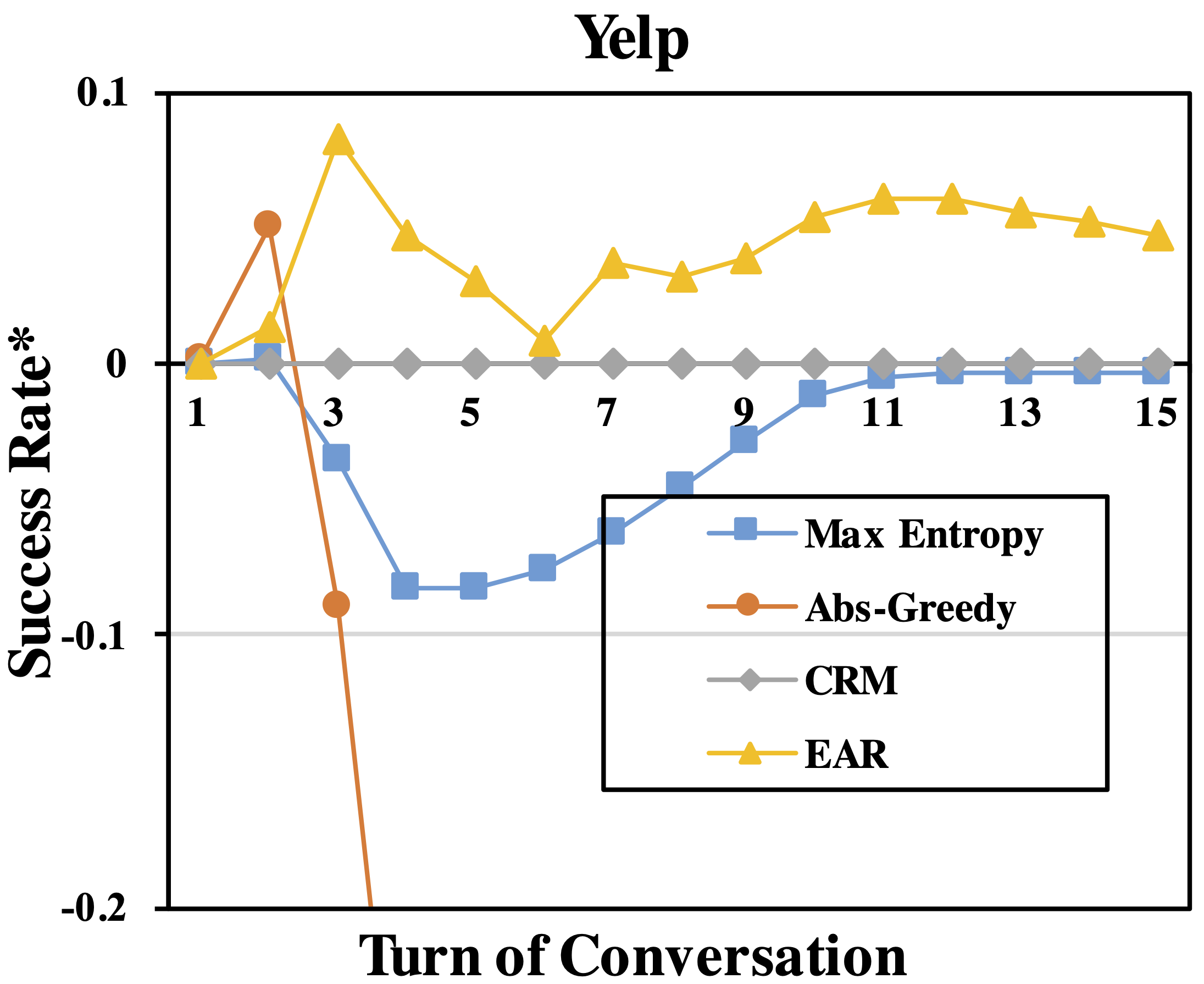}
    \label{f1}
\end{minipage}%
    \hfill%
\begin{minipage}[t]{0.47\linewidth}
    \includegraphics[width=\linewidth]{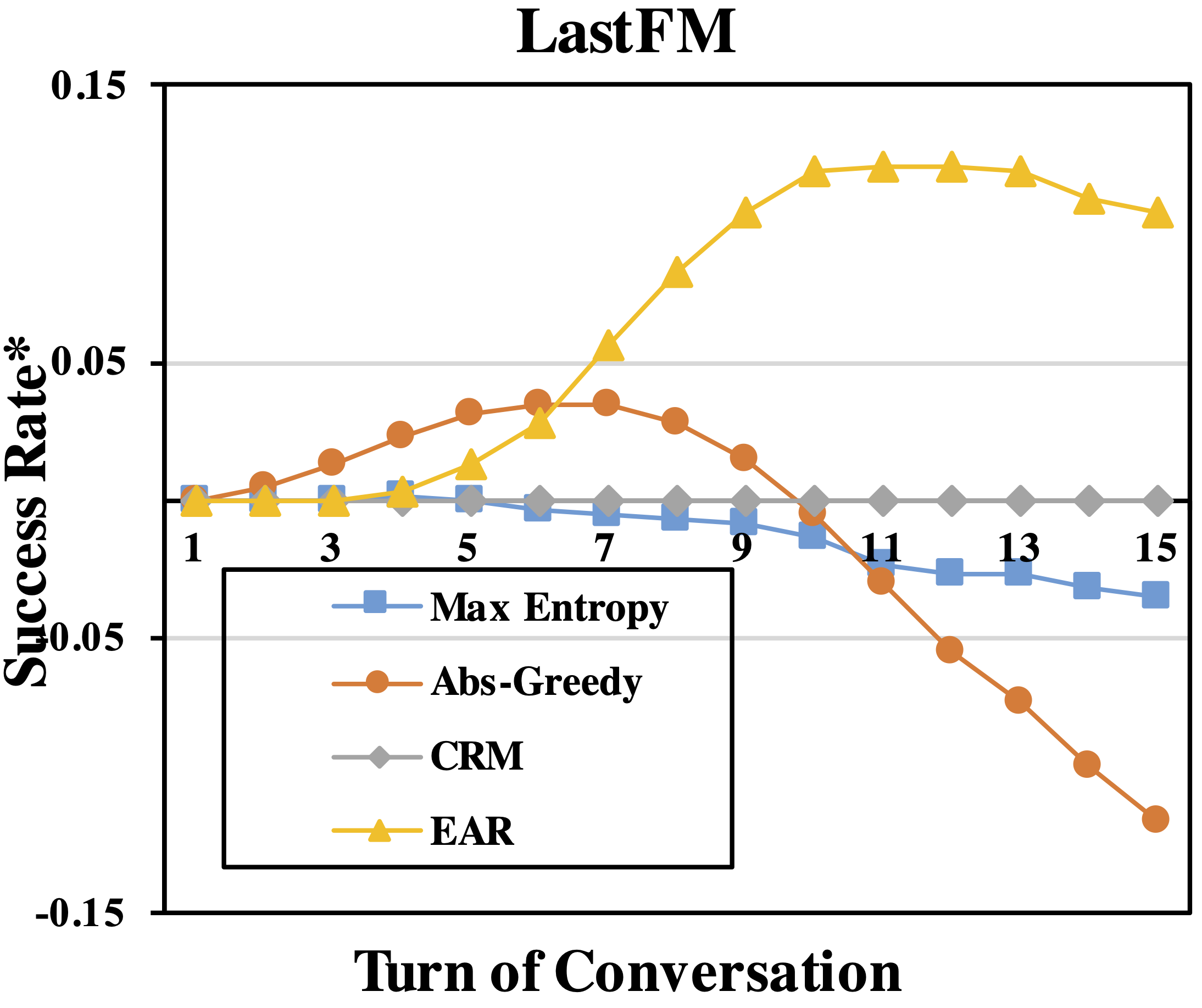}
    \label{f2}
\end{minipage}
\vspace{-15pt}
\caption{Success Rate* of compared methods at different conversation turns on Yelp and LastFM (RQ1).}
\label{fig:overall}
\vspace{-5pt}
\end{figure}

\begin{table}[t]
\caption{SR@15 and AT of compared methods. $^*$ denotes that improvement of EAR over other methods is statistically significant for $p<0.01$ (RQ1).}
\vspace{-5pt}
\label{tab:result}\small
\begin{tabular}{m{2.2cm}|m{1cm}|m{0.8cm}|m{1cm}|m{0.8cm}}\hline
& \multicolumn{2}{c|}{\textbf{LastFM}}  & \multicolumn{2}{c}{\textbf{Yelp}}\\ \hline
 & \textbf{SR@15}    & \textbf{AT} & \textbf{SR@15}    & \textbf{AT} \\ \hline
Abs Greedy &0.209 &13.63 & 0.271 & 12.26  \\ \hline
Max Entropy & 0.290 &13.61 &0.919 &5.77 \\ \hline
CRM & 0.325 & 13.43 & 0.923 & 5.33 \\ \hline
\textbf{EAR}  & \textbf{0.429}* & \textbf{12.45}* & \textbf{0.971}* & \textbf{4.71}*\\ \hline
\end{tabular}
\label{tab:main}
\vspace{-10pt}
\end{table}

Figure~\ref{fig:overall} shows the recommendation Success Rate* (SR*) @t at different turns ($t=1$ to 15), SR* denotes the comparison of each method against the strongest baseline CRM,
indicated as $y=0$
in the figure. Table~\ref{tab:main} shows the scores of the final success rate and the average turns. As can be clearly seen, our EAR model significantly outperforms other methods. This validates our hypothesis that considering extensive interactions between the CC and RC is an effective strategy to build conversational a recommender system. We also make the following observations:

Comparing with Abs Greedy, the three attribute-based methods (EAR, Max Entropy and CRM) have nearly zero success rate at the beginning of a conversation ($t<2$). This is because these methods tend to ask questions at the very beginning. 
As the conversation goes, Abs Greedy (which only recommends items) gradually falls behind the attribute-based methods, demonstrating the efficacy of asking attributes in the conversational recommendation scenario.
Note that Abs Greedy has much weaker performance on Yelp compared to LastFM. The key reason is the setting of Yelp is to ask enumerated question, and user's response with multiple finer-grained attributes sharply shrinks the candidate items.

CRM generally underperforms our EAR methods. One of the key reasons is that its state vector cannot help CC to learn sophisticated strategy to ask and recommend, especially in a much larger action space, i.e., the number of attributes (nearly 30 in our experiments versus 5 in theirs~\cite{Sun:2018:CRS:3209978.3210002}). This result suggests that in a more complex  \textit{multi-round} scenario where the CC needs to make a comprehensive utilization of both the CC (e.g., considering dialogue histories) and RC (considering statistics like attribute preference estimation) when formulating a recommendation strategy. 

Interestingly, 
Figure~\ref{fig:overall} indicates that in Yelp, EAR's gain over CRM enlarges in Turns~1--3, shrinks in Turns~4--6 and widens again afterwards. However, in LastFM it has a steadily increasing gain. This interesting phenomenon reveals that our EAR system can learn different strategies in different settings. In the Yelp dataset, the CRS asks enumerated questions where the user can choose finer-grained attributes, resulting a sharp reduction in the candidate space. The strategy that the EAR system learns is more aggressive: it attempts to ask attributes that can sharply shrink the candidate space and make decisive recommendation at the beginning turns when it feels confident. If this aggressive strategy fails, it changes to a more patient strategy to ask more questions without recommendations, causing less success in the medial turns (e.g., Turns~5--7). However, this strategy pays off in the long term, making recommendation more successful in the latter half of conversations (e.g., after Turn~7). At the same time, CRM is only able to follow the strategy of trying to ask more attributes at the beginning and making recommendations later. In the LastFM dataset, the setting is limited to binary attributes, leading to less efficiency in reducing candidate space. Both EAR and CRM adapt and ask more questions at the outset before making recommendations. However, as EAR incorporates better CC and RC to model better interaction, it significantly outperforms CRM.

\subsection{Effectiveness of Estimation Designs (RQ2)}
\label{sec:main_result}
There are two key designs in the \text{estimation} stage that trains the recommendation model FM offline: the attribute-aware BPR that samples negatives with attribute matching considered, and the multi-task training that jointly optimizes item prediction and attribute prediction tasks. Table \ref{tab:RS_offline} shows offline AUC scores on the two tasks of three methods: FM, FM with attribute-aware BPR (FM+A), and FM+A with multi-task training (FM+A+MT).

\begin{table}[t]
\label{tab:RS_offline}
\small
\caption{Offline AUC score of FM, FM with attribute-aware BPR (FM+A) and with multi-task training for item recommendation and attribute prediction (FM+A+MT). $^*$ denotes that improvement of FM+A+MT over FM and FM+A is statistically significant for $p<0.01$ (RQ2).}
\vspace{-5pt}
\begin{tabular}{m{1.3cm}|>{\centering\arraybackslash}m{1.3cm}|>{\centering\arraybackslash}m{1.3cm}|>{\centering\arraybackslash}m{1.3cm}|>{\centering\arraybackslash}m{1.3cm}}\hline
& \multicolumn{2}{c|}{\textbf{LastFM}}  & \multicolumn{2}{c}{\textbf{Yelp}}\\ \hline
 & \textbf{Item}   & \textbf{Attribute}  & \textbf{Item}  & \textbf{Attribute}  \\ \hline
FM & 0.521 & 0.727 & 0.834 & 0.654\\ \hline
FM+A & 0.724 & 0.629 & 0.866 & 0.638\\ \hline
FM+A+MT & \textbf{0.742}* & \textbf{0.760}* & \textbf{0.870}* & \textbf{0.896}* \\ \hline
\end{tabular}
\vspace{-8pt}
\label{tab:RS_offline}
\end{table}

As can be seen, the attribute-aware BPR significantly boosts the performance of item ranking, being highly beneficial to rank the ground truth item high. Interestingly, it harms the performance of attribute prediction, e.g. on lastFM, FM+A has a much lower AUC score (0.629) than FM (0.727).
The reason might be that the attribute-aware BPR loss guides the model to specifically fit to item ranking in the candidate list. Without an even optimization enforced for the attribute prediction task, it may suffer from poor performance. This implies the necessity of explicitly optimizing the attribute prediction task.
As expected, the best performance is achieved when we add multi-task training on. FM+A+MT significantly enhances the performance of both tasks, validating the effectiveness and rationality of our multi-task training design.

\subsection{Ablation Studies on State Vector (RQ3)}



What information helps in decision making? Let us examine the effects of the the four forms of information included in EAR state vector $\textbf{s}$ (Equation~\ref{eq:state}), by ablating each information type from the feature vector (Table~\ref{ablation}).

\begin{table}[t]
\small
\label{results of relation ablation}\small
\caption{Performance of removing one component of the state vector (Equation~\ref{eq:state}) from our EAR. $^*$ denotes that improvement of EAR over model with removed component is statistically significant for $p<0.01$ (RQ 3).}
\vspace{-5pt}
\begin{tabular}{m{0.7cm}|>{\arraybackslash}m{0.6cm}>{\arraybackslash}m{0.6cm}>{\arraybackslash}m{0.6cm}>{\arraybackslash}m{0.6cm}
|>{\arraybackslash}m{0.6cm}>{\arraybackslash}m{0.6cm}>{\arraybackslash}m{0.6cm}>{\arraybackslash}m{0.6cm}}\hline
& \multicolumn{4}{c|}{\textbf{Yelp}}  & \multicolumn{4}{c}{\textbf{LastFM}}\\ \hline
 & SR@5   & SR@10  & SR@15  & AT & SR@5   & SR@10 & SR@15  & AT\\ \hline
$-\textbf{s}_{ent}$ &0.614&0.895&0.969&4.81 &\bf 0.051&0.190&0.346&12.82 \\ \hline
$-\textbf{s}_{pre}$ &0.596&0.857&0.959&5.06&0.024&0.231&0.407&12.55\\ \hline
$-\textbf{s}_{his}$ &0.624&0.894&0.949&4.79&0.021&0.236&0.424&12.50 \\ \hline
$-\textbf{s}_{len}$ &0.550&0.846&0.952&5.44&0.013&0.230&0.416&12.56 \\ \hline
\textbf{EAR} &\bf0.629*&\bf0.907*&\bf0.971*&\bf4.71* & 0.020&\bf 0.243*&\bf0.429*&\bf12.45*\\ \hline
\end{tabular}
\label{ablation}
\vspace{-10pt}
\end{table}

Comparing the performance drop of each method, we uncover differences that corroborate the intrinsic difference between the two conversational settings. The most important factor is question type: i.e., $\textbf{s}_{ent}$ for LastFM (binary question) and $\textbf{s}_{len}$ for Yelp (enumerated question). The entropy($\textbf{s}_{ent}$) information is crucial for LastFM, it is in line with the claim in ~\cite{dhingra2017towards} that the maximum entropy is the best strategy when language understanding is precise. 
If we ablate $\textbf{s}_{ent}$ on LastFM, although it reaches 0.051 in SR@5, future SR greatly suffers, due to the system's over-agressiveness to recommend items before obtaining sufficient relevant attribute evidence.
As for the enumerated question setting (Yelp), the candidate list length ($\textbf{s}_{len}$) is most important, because the candidate item list shrinks more sharply and $\textbf{s}_{len}$ is helpful when deciding when to recommend.

Apart from entropy and candidate list length, the remaining two factors -- i.e., attribute preference, conversation history -- both contribute positively.
Their impact is sensitive to datasets and metrics. For example, the attribute preference ($\textbf{s}_{pre}$) strongly affect SR@5 and SR@10 on Yelp, but does not show significant impacts for SR@15.
This inconsistency provides an evidence for the intrinsic difficulty of decision making in the conversational recommendation scenario, which however has yet to be extensively studied.
\subsection{Investigation on Reflection (RQ4)}

\begin{table}[t]
\small
\caption{Performance after removing the online update module in the reflection stage. $^*$ denotes that improvement of EAR over removing update module is statistically significant for $p<0.01$ (RQ4).}\label{tab:RQ4}
\vspace{-5pt}
\begin{tabular}{m{1cm}|>{\arraybackslash}m{0.6cm}>{\arraybackslash}m{0.6cm}>{\arraybackslash}m{0.6cm}>{\arraybackslash}m{0.6cm}
|>{\arraybackslash}m{0.6cm}>{\arraybackslash}m{0.6cm}>{\arraybackslash}m{0.6cm}>{\arraybackslash}m{0.6cm}}\hline
& \multicolumn{4}{c|}{\textbf{Yelp}}  & \multicolumn{4}{c}{\textbf{LastFM}}\\ \hline
 & SR@5   & SR@10  & SR@15  & AT & SR@5   & SR@10 & SR@15  & AT\\ \hline
-\textbf{update} &0.629& 0.905 & 0.970 & 4.72 &0.020& 0.217 & 0.393 & 12.67 \\ \hline
\textbf{EAR} &\bf0.629&\bf0.907&\bf0.971&\bf4.71 & \bf0.020&\bf 0.243*&\bf0.429*&\bf12.45*\\ \hline
\end{tabular}
\vspace{-10pt}
\end{table}

\begin{figure}[t]
    \centering
    \subfigure{
	\includegraphics[height=3.6cm]{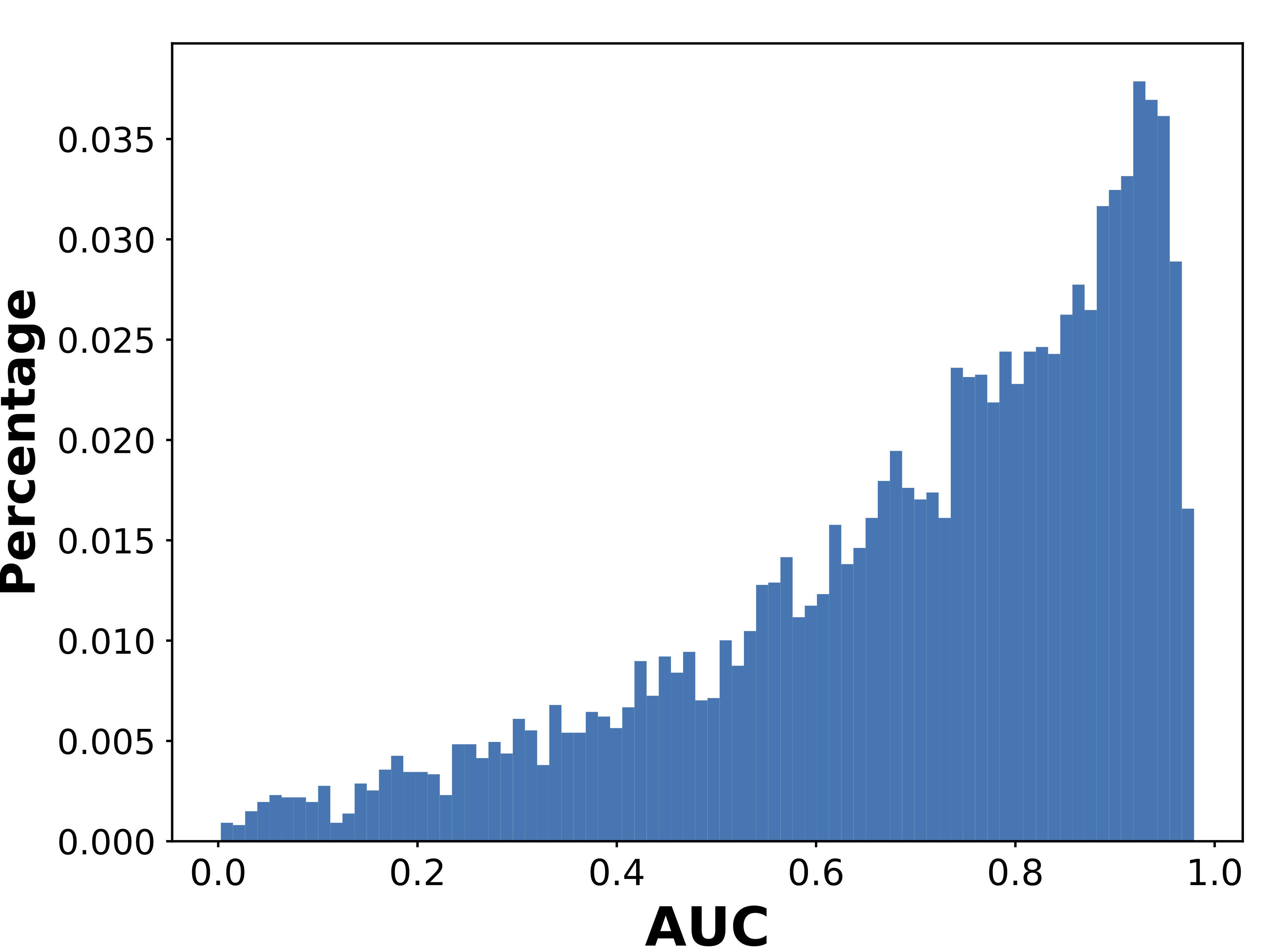}}\vspace{-5pt}
    \caption{Percentage of bad updates w.r.t. the offline model's AUC on the users on Yelp (RQ4).
    }\label{fig:update}
    \vspace{-10pt}
\end{figure}

\begin{table*}[t]\centering
\caption{Recent conversational recommender summary: 1) whether it asks about attributes, 2) question space, 3) any explicit strategy w.r.t. recommendation timing, 4) whether it considers multi-round recommendations, and 5) its main focus.}\label{tab:CRS}\vspace{-8pt}
  \small
  \begin{tabular}{m{2.8cm}|>{\centering\arraybackslash}m{0.8cm}|>{\centering\arraybackslash}m{2.4cm}|>{\centering\arraybackslash}m{1.4cm}|>{\centering\arraybackslash}m{2cm}|m{6.5cm}}\hline
                 & \textbf{1. Q?}  & \textbf{2. Question Space} & \textbf{3. Explicit} & \textbf{4. Multi-round} & \textbf{5. Main Focus} \\ \hline
Online bandits~\cite{christakopoulou2016towards,wu2016contextual,wu2018learning}& $\times$ & N.A. & $\times$ & $\checkmark$ &   Exploration-exploitation trade-off in item selection \\ \hline
REDIAL (NIPS'18)~\cite{nips18/DeepConv} &  $\checkmark$ & Free texts & $\times$ &$\checkmark$ & End-to-end generation of natural language response \\ \hline
KMD (MM'18)~\cite{Liao:2018} & $\checkmark$ & Free texts & $\times$ & $\checkmark$ & End-to-end generation of text and image response \\ \hline
Q\&R (KDD'18)~\cite{christakopoulou2018q}        & $\checkmark$  & Attributes & $\times$ &  $\times$  & Question asking and single-round recommendation \\ \hline
MMN (CIKM'18)~\cite{zhang2018towards} &  $\checkmark$  &  Attributes  &  $\times$  & $\checkmark$  & Attribute-product match in conversational search \\ \hline
CRM (SIGIR'18)~\cite{Sun:2018:CRS:3209978.3210002}       & $\checkmark$   & Attributes  &  $\checkmark$    &    $\times$   &  Shallow combination between CC and RC  \\ \hline
VDARIS (KDD'19)~\cite{yu2019visual}       & $\times$   & N.A.  &  $\times$    &    $\checkmark$   &  User's click and comment on recommended items \\ \hline
EAR (our method) & $\checkmark$ & Attributes &$\checkmark$ &$\checkmark$ & Deep interaction between CC and RC  \\ \hline
\end{tabular}
\vspace{-5pt}
\end{table*}

To understand the impact of online update in the reflection stage, we start from the ablation study.  Table~\ref{tab:RQ4} shows the variant of EAR that removes online update. We find that the trends do not converge on two datasets: the updating strategy helps a lot on LastFM but has very minor effect on the Yelp dataset.

Questioning this interesting phenomenon, we examine the individual items on Yelp.
We find that the updating does not always help ranking, especially when the offline model already ranks the ground truth item high (but not at top 10).
In this case, doing updates is highly likely to pull down the ranking position of the ground truth item.
To gain statistical evidence for this observation,
we term such updates as \textit{bad updates}, and show the percentage of bad updates with respect to the offline model's AUC on the users. As seen from Figure~\ref{fig:update}, there is a clear positive correlation between bad updates and AUC score. For example, $\sim$3.5\% of the bad updates come from users with an offline AUC of 0.9.

This explains why online update works well for LastFM, but not for Yelp:
our recommendation model has a better performance on Yelp than LastFM (0.870 v.s. 0.742 in AUC as shown in Table~\ref{tab:RS_offline}).
This means the items on Yelp are more likely to get higher AUC, resulting in worse updates.
More such observations and analyses will help further the community understanding the efficacy of online updates.
Although bandit algorithms have devoted to exploring this question~\cite{kuleshov2014algorithms,chu2011contextual,li2016collaborative,gentile2014online,wu2016contextual}, the issue has largely been unaddressed in the context of conversational recommender system.

\section{Related Work}\label{sec:related}
The offline \textbf{static recommendation} task is formulated as estimating the affinity score between a user and an item~\cite{NCF}. This is usually achieved by learning user preferences through the historical user-item interactions such as clicking and purchasing. The representative methods are Matrix Factorization (MF)~\cite{MF} and Factorization Machine (FM)~\cite{rendle2010factorization}. Neural FM~\cite{NFM} and DeepFM~\cite{guo2017deepfm} have improved FM's representation ability with deep neural networks. \cite{fastMF,iCD,sigir/EbesuSF18} utilize user's implicit feedback, commonly optimizing BPR loss~\cite{BPR}. \cite{cheng2019mmalfm, cheng2018aspect} exploits user's reviews and image information.  However, such static recommendation methods suffer from the intrinsic limitation of not being able to capture user dynamic preferences.

This intrinsic limitation motivates \textbf{online recommendation}. Its target is to adapt the recommendation results with the user's online actions~\cite{li2015online}. Many model it as a multi-arm bandit problem
~\cite{wu2016contextual,wang2017factorization,wu2018learning}
, strategically demonstrating items to users for useful feedback. \cite{zhang2019toward} makes the preliminary effort to extend the bandit framework to query attributes. While achieving remarkable progress, the bandit-based solutions are still insufficient: 1) Such methods focus on exploration--exploitation trade-off in cold start settings. However, in warm start scenario, capturing the user dynamic preference is critical as preference drift is common; 2) The mathematical formation of multi-arm bandit problem limits such method only recommend one item each time. This constraint limits its application, as we usually need to recommend a list of items.

\textbf{Conversational recommender systems} provide a new possibility for capturing dynamic feedback as they enable a system to interact with users using natural language. However, they also pose challenges to researchers, leading to various settings and problem formulations ~\cite{christakopoulou2016towards,nips18/DeepConv,Liao:2018, christakopoulou2018q,zhang2018towards,Sun:2018:CRS:3209978.3210002, priyogi2019preference, liao2019deep, yu2019visual, ayundhita2019ontology, sardella2019approach, zhang2019toward}. Table~\ref{tab:CRS}  summarizes these works' key aspects. Generally, prior work considers conversational recommendation only under simplified settings. For example, \cite{christakopoulou2016towards, yu2019visual} only allow the CRS to recommend items without asking the user about their preferred attributes. The Q\&R work~\cite{christakopoulou2018q} proposes to jointly optimize the two tasks of attribute and item prediction, but restricts the whole conversation to two turns: one turn for asking, one turn for  recommending.  CRM~\cite{Sun:2018:CRS:3209978.3210002} extends the conversation to multi-turns but still follows the single-round setting. MMN~\cite{zhang2018towards} focuses on document representation, aiming to learn better matching function for attributes and products description under a conversation setting. Unfortunately, it does not build a dialogue policy to decide when to ask or make recommendations. 
In contrast, situations for various real applications are complex: the CRS needs to strategically ask attributes and make recommendations in multiple rounds, achieving successful recommendations in the fewest turns. In recent work, only~\cite{nips18/DeepConv} considers this multi-round scenario, but it focuses on language understanding and generation, without attending to explicitly model the conversational strategy.

\section{Conclusion and Future Work}
In this work, we redefine the conversational recommendation task where the RC and CC closely support each other so as to achieve the goal of accurate recommendation in fewer turns. 
We decompose the task into three key problems, namely, what to ask, when to recommend, and how to adapt with user feedback. We then propose EAR -- a new three-stage solution accounting for the three problems in a unified framework. For each stage, we design our method to carefully account for the interactions between RC and CC. Through extensive experiments on two datasets, we justify the effectiveness of EAR, providing additional insights into the conversational strategy and online updates. 




Our work represents the first step towards exploring how the CC and RC can collaborate closely to provide quality recommendation service in this multi-round scenario. Naturally, there are thus a few loose ends for further investigation, especially with respect to incorporating user feedback. 
In the future, we will consider refreshing the policy network to make better actions. 
We will also extend EAR to consider explore--exploit balance which is the key problem for traditional interactive recommendation system.
Lastly, we will deploy our system to online applications that interact with real users to gain more insights for further improvements. 

\textbf{Acknowledgement}:
This research is part of NExT++ research and also supported by the National Natural Science Foundation of China (61972372).
NExT++ is supported by the National Research Foundation, Prime Minister's Office, Singapore under its IRC@SG Funding Initiative.
We would like to thank the anonymous reviewers for their valuable reviews. 

\vspace{-10pt}



\bibliographystyle{ACM-Reference-Format}
\bibliography{reference.bib}


\begin{thebibliography}{40}


\ifx \showCODEN    \undefined \def \showCODEN     #1{\unskip}     \fi
\ifx \showDOI      \undefined \def \showDOI       #1{#1}\fi
\ifx \showISBNx    \undefined \def \showISBNx     #1{\unskip}     \fi
\ifx \showISBNxiii \undefined \def \showISBNxiii  #1{\unskip}     \fi
\ifx \showISSN     \undefined \def \showISSN      #1{\unskip}     \fi
\ifx \showLCCN     \undefined \def \showLCCN      #1{\unskip}     \fi
\ifx \shownote     \undefined \def \shownote      #1{#1}          \fi
\ifx \showarticletitle \undefined \def \showarticletitle #1{#1}   \fi
\ifx \showURL      \undefined \def \showURL       {\relax}        \fi
\providecommand\bibfield[2]{#2}
\providecommand\bibinfo[2]{#2}
\providecommand\natexlab[1]{#1}
\providecommand\showeprint[2][]{arXiv:#2}

\bibitem[\protect\citeauthoryear{Auer}{Auer}{2002}]%
        {auer2002using}
\bibfield{author}{\bibinfo{person}{Peter Auer}.}
  \bibinfo{year}{2002}\natexlab{}.
\newblock \showarticletitle{Using confidence bounds for
  exploitation-exploration trade-offs}.
\newblock \bibinfo{journal}{\emph{Journal of Machine Learning Research}}
  \bibinfo{volume}{3}, \bibinfo{number}{Nov} (\bibinfo{year}{2002}),
  \bibinfo{pages}{397--422}.
\newblock


\bibitem[\protect\citeauthoryear{Ayundhita, Baizal, and Sibaroni}{Ayundhita
  et~al\mbox{.}}{2019}]%
        {ayundhita2019ontology}
\bibfield{author}{\bibinfo{person}{MS Ayundhita}, \bibinfo{person}{ZKA Baizal},
  {and} \bibinfo{person}{Y Sibaroni}.} \bibinfo{year}{2019}\natexlab{}.
\newblock \showarticletitle{Ontology-based conversational recommender system
  for recommending laptop}. In \bibinfo{booktitle}{\emph{Journal of Physics:
  Conference Series}}, Vol.~\bibinfo{volume}{1192}. IOP Publishing,
  \bibinfo{pages}{012020}.
\newblock


\bibitem[\protect\citeauthoryear{Bayer, He, Kanagal, and Rendle}{Bayer
  et~al\mbox{.}}{2017}]%
        {iCD}
\bibfield{author}{\bibinfo{person}{Immanuel Bayer}, \bibinfo{person}{Xiangnan
  He}, \bibinfo{person}{Bhargav Kanagal}, {and} \bibinfo{person}{Steffen
  Rendle}.} \bibinfo{year}{2017}\natexlab{}.
\newblock \showarticletitle{A generic coordinate descent framework for learning
  from implicit feedback}. In \bibinfo{booktitle}{\emph{WWW}}.
  \bibinfo{pages}{1341--1350}.
\newblock


\bibitem[\protect\citeauthoryear{Chapelle and Li}{Chapelle and Li}{2011}]%
        {chapelle2011empirical}
\bibfield{author}{\bibinfo{person}{Olivier Chapelle} {and}
  \bibinfo{person}{Lihong Li}.} \bibinfo{year}{2011}\natexlab{}.
\newblock \showarticletitle{An empirical evaluation of thompson sampling}. In
  \bibinfo{booktitle}{\emph{NeurIPS}}. \bibinfo{pages}{2249--2257}.
\newblock


\bibitem[\protect\citeauthoryear{Chen, Ren, Tang, Zhao, and Yin}{Chen
  et~al\mbox{.}}{2018}]%
        {chen2018hierarchical}
\bibfield{author}{\bibinfo{person}{Hongshen Chen}, \bibinfo{person}{Zhaochun
  Ren}, \bibinfo{person}{Jiliang Tang}, \bibinfo{person}{Yihong~Eric Zhao},
  {and} \bibinfo{person}{Dawei Yin}.} \bibinfo{year}{2018}\natexlab{}.
\newblock \showarticletitle{Hierarchical Variational Memory Network for
  Dialogue Generation}. In \bibinfo{booktitle}{\emph{WWW}}.
  \bibinfo{pages}{1653--1662}.
\newblock


\bibitem[\protect\citeauthoryear{Chen, Zhang, He, Nie, Liu, and Chua}{Chen
  et~al\mbox{.}}{2017}]%
        {ACF}
\bibfield{author}{\bibinfo{person}{Jingyuan Chen}, \bibinfo{person}{Hanwang
  Zhang}, \bibinfo{person}{Xiangnan He}, \bibinfo{person}{Liqiang Nie},
  \bibinfo{person}{Wei Liu}, {and} \bibinfo{person}{Tat{-}Seng Chua}.}
  \bibinfo{year}{2017}\natexlab{}.
\newblock \showarticletitle{Attentive Collaborative Filtering: Multimedia
  Recommendation with Item- and Component-Level Attention}. In
  \bibinfo{booktitle}{\emph{SIGIR}}. \bibinfo{pages}{335--344}.
\newblock


\bibitem[\protect\citeauthoryear{Cheng, Chang, Zhu, Kanjirathinkal, and
  Kankanhalli}{Cheng et~al\mbox{.}}{2019}]%
        {cheng2019mmalfm}
\bibfield{author}{\bibinfo{person}{Zhiyong Cheng}, \bibinfo{person}{Xiaojun
  Chang}, \bibinfo{person}{Lei Zhu}, \bibinfo{person}{Rose~C Kanjirathinkal},
  {and} \bibinfo{person}{Mohan Kankanhalli}.} \bibinfo{year}{2019}\natexlab{}.
\newblock \showarticletitle{MMALFM: Explainable recommendation by leveraging
  reviews and images}.
\newblock \bibinfo{journal}{\emph{TOIS}} \bibinfo{volume}{37},
  \bibinfo{number}{2} (\bibinfo{year}{2019}), \bibinfo{pages}{16}.
\newblock


\bibitem[\protect\citeauthoryear{Cheng, Ding, Zhu, and Kankanhalli}{Cheng
  et~al\mbox{.}}{2018}]%
        {cheng2018aspect}
\bibfield{author}{\bibinfo{person}{Zhiyong Cheng}, \bibinfo{person}{Ying Ding},
  \bibinfo{person}{Lei Zhu}, {and} \bibinfo{person}{Mohan Kankanhalli}.}
  \bibinfo{year}{2018}\natexlab{}.
\newblock \showarticletitle{Aspect-aware latent factor model: Rating prediction
  with ratings and reviews}. In \bibinfo{booktitle}{\emph{Proceedings of the
  2018 World Wide Web Conference}}. International World Wide Web Conferences
  Steering Committee, \bibinfo{pages}{639--648}.
\newblock


\bibitem[\protect\citeauthoryear{Christakopoulou, Beutel, Li, Jain, and
  Chi}{Christakopoulou et~al\mbox{.}}{2018}]%
        {christakopoulou2018q}
\bibfield{author}{\bibinfo{person}{Konstantina Christakopoulou},
  \bibinfo{person}{Alex Beutel}, \bibinfo{person}{Rui Li},
  \bibinfo{person}{Sagar Jain}, {and} \bibinfo{person}{Ed~H Chi}.}
  \bibinfo{year}{2018}\natexlab{}.
\newblock \showarticletitle{Q\&R: A Two-Stage Approach toward Interactive
  Recommendation}. In \bibinfo{booktitle}{\emph{SIGKDD}}.
  \bibinfo{pages}{139--148}.
\newblock


\bibitem[\protect\citeauthoryear{Christakopoulou, Radlinski, and
  Hofmann}{Christakopoulou et~al\mbox{.}}{2016}]%
        {christakopoulou2016towards}
\bibfield{author}{\bibinfo{person}{Konstantina Christakopoulou},
  \bibinfo{person}{Filip Radlinski}, {and} \bibinfo{person}{Katja Hofmann}.}
  \bibinfo{year}{2016}\natexlab{}.
\newblock \showarticletitle{Towards conversational recommender systems}. In
  \bibinfo{booktitle}{\emph{SIGKDD}}. \bibinfo{pages}{815--824}.
\newblock


\bibitem[\protect\citeauthoryear{Chu, Li, Reyzin, and Schapire}{Chu
  et~al\mbox{.}}{2011}]%
        {chu2011contextual}
\bibfield{author}{\bibinfo{person}{Wei Chu}, \bibinfo{person}{Lihong Li},
  \bibinfo{person}{Lev Reyzin}, {and} \bibinfo{person}{Robert Schapire}.}
  \bibinfo{year}{2011}\natexlab{}.
\newblock \showarticletitle{Contextual bandits with linear payoff functions}.
  In \bibinfo{booktitle}{\emph{Proceedings of the Fourteenth International
  Conference on Artificial Intelligence and Statistics}}.
  \bibinfo{pages}{208--214}.
\newblock


\bibitem[\protect\citeauthoryear{Dhingra, Li, Li, Gao, Chen, Ahmed, and
  Deng}{Dhingra et~al\mbox{.}}{2017}]%
        {dhingra2017towards}
\bibfield{author}{\bibinfo{person}{Bhuwan Dhingra}, \bibinfo{person}{Lihong
  Li}, \bibinfo{person}{Xiujun Li}, \bibinfo{person}{Jianfeng Gao},
  \bibinfo{person}{Yun-Nung Chen}, \bibinfo{person}{Faisal Ahmed}, {and}
  \bibinfo{person}{Li Deng}.} \bibinfo{year}{2017}\natexlab{}.
\newblock \showarticletitle{Towards End-to-End Reinforcement Learning of
  Dialogue Agents for Information Access}. In \bibinfo{booktitle}{\emph{ACL}}.
  \bibinfo{pages}{484--495}.
\newblock


\bibitem[\protect\citeauthoryear{Ebesu, Shen, and Fang}{Ebesu
  et~al\mbox{.}}{2018}]%
        {sigir/EbesuSF18}
\bibfield{author}{\bibinfo{person}{Travis Ebesu}, \bibinfo{person}{Bin Shen},
  {and} \bibinfo{person}{Yi Fang}.} \bibinfo{year}{2018}\natexlab{}.
\newblock \showarticletitle{Collaborative Memory Network for Recommendation
  Systems}. In \bibinfo{booktitle}{\emph{SIGIR}}. \bibinfo{pages}{515--524}.
\newblock


\bibitem[\protect\citeauthoryear{Gentile, Li, and Zappella}{Gentile
  et~al\mbox{.}}{2014}]%
        {gentile2014online}
\bibfield{author}{\bibinfo{person}{Claudio Gentile}, \bibinfo{person}{Shuai
  Li}, {and} \bibinfo{person}{Giovanni Zappella}.}
  \bibinfo{year}{2014}\natexlab{}.
\newblock \showarticletitle{Online clustering of bandits}. In
  \bibinfo{booktitle}{\emph{ICML}}. \bibinfo{pages}{757--765}.
\newblock


\bibitem[\protect\citeauthoryear{Guo, Tang, Ye, Li, and He}{Guo
  et~al\mbox{.}}{2017}]%
        {guo2017deepfm}
\bibfield{author}{\bibinfo{person}{Huifeng Guo}, \bibinfo{person}{Ruiming
  Tang}, \bibinfo{person}{Yunming Ye}, \bibinfo{person}{Zhenguo Li}, {and}
  \bibinfo{person}{Xiuqiang He}.} \bibinfo{year}{2017}\natexlab{}.
\newblock \showarticletitle{Deepfm: a factorization-machine based neural
  network for ctr prediction}. In \bibinfo{booktitle}{\emph{IJCAI}}.
\newblock


\bibitem[\protect\citeauthoryear{He and Chua}{He and Chua}{2017}]%
        {NFM}
\bibfield{author}{\bibinfo{person}{Xiangnan He} {and} \bibinfo{person}{Tat-Seng
  Chua}.} \bibinfo{year}{2017}\natexlab{}.
\newblock \showarticletitle{Neural factorization machines for sparse predictive
  analytics}. In \bibinfo{booktitle}{\emph{SIGIR}}. \bibinfo{pages}{355--364}.
\newblock


\bibitem[\protect\citeauthoryear{He, Liao, Zhang, Nie, Hu, and Chua}{He
  et~al\mbox{.}}{2017}]%
        {NCF}
\bibfield{author}{\bibinfo{person}{Xiangnan He}, \bibinfo{person}{Lizi Liao},
  \bibinfo{person}{Hanwang Zhang}, \bibinfo{person}{Liqiang Nie},
  \bibinfo{person}{Xia Hu}, {and} \bibinfo{person}{Tat{-}Seng Chua}.}
  \bibinfo{year}{2017}\natexlab{}.
\newblock \showarticletitle{Neural Collaborative Filtering}. In
  \bibinfo{booktitle}{\emph{WWW}}. \bibinfo{pages}{173--182}.
\newblock


\bibitem[\protect\citeauthoryear{He, Zhang, Kan, and Chua}{He
  et~al\mbox{.}}{2016}]%
        {fastMF}
\bibfield{author}{\bibinfo{person}{Xiangnan He}, \bibinfo{person}{Hanwang
  Zhang}, \bibinfo{person}{Min-Yen Kan}, {and} \bibinfo{person}{Tat-Seng
  Chua}.} \bibinfo{year}{2016}\natexlab{}.
\newblock \showarticletitle{Fast matrix factorization for online recommendation
  with implicit feedback}. In \bibinfo{booktitle}{\emph{SIGIR}}.
  \bibinfo{pages}{549--558}.
\newblock


\bibitem[\protect\citeauthoryear{Jin, Lei, Ren, Chen, Liang, Zhao, and Yin}{Jin
  et~al\mbox{.}}{2018}]%
        {jin2018explicit}
\bibfield{author}{\bibinfo{person}{Xisen Jin}, \bibinfo{person}{Wenqiang Lei},
  \bibinfo{person}{Zhaochun Ren}, \bibinfo{person}{Hongshen Chen},
  \bibinfo{person}{Shangsong Liang}, \bibinfo{person}{Yihong Zhao}, {and}
  \bibinfo{person}{Dawei Yin}.} \bibinfo{year}{2018}\natexlab{}.
\newblock \showarticletitle{Explicit State Tracking with Semi-Supervisionfor
  Neural Dialogue Generation}. In \bibinfo{booktitle}{\emph{CIKM}}. ACM,
  \bibinfo{pages}{1403--1412}.
\newblock


\bibitem[\protect\citeauthoryear{Koren, Bell, and Volinsky}{Koren
  et~al\mbox{.}}{2009}]%
        {MF}
\bibfield{author}{\bibinfo{person}{Yehuda Koren}, \bibinfo{person}{Robert~M.
  Bell}, {and} \bibinfo{person}{Chris Volinsky}.}
  \bibinfo{year}{2009}\natexlab{}.
\newblock \showarticletitle{Matrix Factorization Techniques for Recommender
  Systems}.
\newblock \bibinfo{journal}{\emph{{IEEE} Computer}} \bibinfo{volume}{42},
  \bibinfo{number}{8} (\bibinfo{year}{2009}), \bibinfo{pages}{30--37}.
\newblock


\bibitem[\protect\citeauthoryear{Kuleshov and Precup}{Kuleshov and
  Precup}{2014}]%
        {kuleshov2014algorithms}
\bibfield{author}{\bibinfo{person}{Volodymyr Kuleshov} {and}
  \bibinfo{person}{Doina Precup}.} \bibinfo{year}{2014}\natexlab{}.
\newblock \showarticletitle{Algorithms for multi-armed bandit problems}.
\newblock \bibinfo{journal}{\emph{arXiv preprint arXiv:1402.6028}}
  (\bibinfo{year}{2014}).
\newblock


\bibitem[\protect\citeauthoryear{Lei, Jin, Kan, Ren, He, and Yin}{Lei
  et~al\mbox{.}}{2018}]%
        {acl18/sequicity}
\bibfield{author}{\bibinfo{person}{Wenqiang Lei}, \bibinfo{person}{Xisen Jin},
  \bibinfo{person}{Min{-}Yen Kan}, \bibinfo{person}{Zhaochun Ren},
  \bibinfo{person}{Xiangnan He}, {and} \bibinfo{person}{Dawei Yin}.}
  \bibinfo{year}{2018}\natexlab{}.
\newblock \showarticletitle{Sequicity: Simplifying Task-oriented Dialogue
  Systems with Single Sequence-to-Sequence Architectures}. In
  \bibinfo{booktitle}{\emph{ACL}}. \bibinfo{pages}{1437--1447}.
\newblock


\bibitem[\protect\citeauthoryear{Li, Kahou, Schulz, Michalski, Charlin, and
  Pal}{Li et~al\mbox{.}}{2018}]%
        {nips18/DeepConv}
\bibfield{author}{\bibinfo{person}{Raymond Li},
  \bibinfo{person}{Samira~Ebrahimi Kahou}, \bibinfo{person}{Hannes Schulz},
  \bibinfo{person}{Vincent Michalski}, \bibinfo{person}{Laurent Charlin}, {and}
  \bibinfo{person}{Chris Pal}.} \bibinfo{year}{2018}\natexlab{}.
\newblock \showarticletitle{Towards Deep Conversational Recommendations}. In
  \bibinfo{booktitle}{\emph{NeurIPS}}. \bibinfo{pages}{9748--9758}.
\newblock


\bibitem[\protect\citeauthoryear{Li, Karatzoglou, and Gentile}{Li
  et~al\mbox{.}}{2016}]%
        {li2016collaborative}
\bibfield{author}{\bibinfo{person}{Shuai Li}, \bibinfo{person}{Alexandros
  Karatzoglou}, {and} \bibinfo{person}{Claudio Gentile}.}
  \bibinfo{year}{2016}\natexlab{}.
\newblock \showarticletitle{Collaborative filtering bandits}. In
  \bibinfo{booktitle}{\emph{SIGIR}}. \bibinfo{pages}{539--548}.
\newblock


\bibitem[\protect\citeauthoryear{Li and Karahanna}{Li and Karahanna}{2015}]%
        {li2015online}
\bibfield{author}{\bibinfo{person}{Seth~Siyuan Li} {and} \bibinfo{person}{Elena
  Karahanna}.} \bibinfo{year}{2015}\natexlab{}.
\newblock \showarticletitle{Online recommendation systems in a B2C E-commerce
  context: a review and future directions}.
\newblock \bibinfo{journal}{\emph{Journal of the Association for Information
  Systems}} \bibinfo{volume}{16}, \bibinfo{number}{2} (\bibinfo{year}{2015}),
  \bibinfo{pages}{72}.
\newblock


\bibitem[\protect\citeauthoryear{Liao, Ma, He, Hong, and Chua}{Liao
  et~al\mbox{.}}{2018}]%
        {Liao:2018}
\bibfield{author}{\bibinfo{person}{Lizi Liao}, \bibinfo{person}{Yunshan Ma},
  \bibinfo{person}{Xiangnan He}, \bibinfo{person}{Richang Hong}, {and}
  \bibinfo{person}{Tat-Seng Chua}.} \bibinfo{year}{2018}\natexlab{}.
\newblock \showarticletitle{Knowledge-aware Multimodal Dialogue Systems}. In
  \bibinfo{booktitle}{\emph{ACM MM}}. \bibinfo{pages}{801--809}.
\newblock


\bibitem[\protect\citeauthoryear{Liao, Takanobu, Ma, Yang, Huang, and
  Chua}{Liao et~al\mbox{.}}{2019}]%
        {liao2019deep}
\bibfield{author}{\bibinfo{person}{Lizi Liao}, \bibinfo{person}{Ryuichi
  Takanobu}, \bibinfo{person}{Yunshan Ma}, \bibinfo{person}{Xun Yang},
  \bibinfo{person}{Minlie Huang}, {and} \bibinfo{person}{Tat-Seng Chua}.}
  \bibinfo{year}{2019}\natexlab{}.
\newblock \showarticletitle{Deep Conversational Recommender in Travel}.
\newblock \bibinfo{journal}{\emph{arXiv preprint arXiv:1907.00710}}
  (\bibinfo{year}{2019}).
\newblock


\bibitem[\protect\citeauthoryear{Priyogi}{Priyogi}{2019}]%
        {priyogi2019preference}
\bibfield{author}{\bibinfo{person}{Bilih Priyogi}.}
  \bibinfo{year}{2019}\natexlab{}.
\newblock \showarticletitle{Preference Elicitation Strategy for Conversational
  Recommender System}. In \bibinfo{booktitle}{\emph{WSDM}}. ACM,
  \bibinfo{pages}{824--825}.
\newblock


\bibitem[\protect\citeauthoryear{Rendle}{Rendle}{2010}]%
        {rendle2010factorization}
\bibfield{author}{\bibinfo{person}{Steffen Rendle}.}
  \bibinfo{year}{2010}\natexlab{}.
\newblock \showarticletitle{Factorization machines}. In
  \bibinfo{booktitle}{\emph{ICDM}}. IEEE, \bibinfo{pages}{995--1000}.
\newblock


\bibitem[\protect\citeauthoryear{Rendle, Freudenthaler, Gantner, and
  Schmidt-Thieme}{Rendle et~al\mbox{.}}{2009}]%
        {BPR}
\bibfield{author}{\bibinfo{person}{Steffen Rendle}, \bibinfo{person}{Christoph
  Freudenthaler}, \bibinfo{person}{Zeno Gantner}, {and} \bibinfo{person}{Lars
  Schmidt-Thieme}.} \bibinfo{year}{2009}\natexlab{}.
\newblock \showarticletitle{BPR: Bayesian personalized ranking from implicit
  feedback}. In \bibinfo{booktitle}{\emph{UAI}}.
\newblock


\bibitem[\protect\citeauthoryear{Sardella, Biancalana, Micarelli, and
  Sansonetti}{Sardella et~al\mbox{.}}{2019}]%
        {sardella2019approach}
\bibfield{author}{\bibinfo{person}{Nicola Sardella}, \bibinfo{person}{Claudio
  Biancalana}, \bibinfo{person}{Alessandro Micarelli}, {and}
  \bibinfo{person}{Giuseppe Sansonetti}.} \bibinfo{year}{2019}\natexlab{}.
\newblock \showarticletitle{An Approach to Conversational Recommendation of
  Restaurants}. In \bibinfo{booktitle}{\emph{ICHCI}}. Springer,
  \bibinfo{pages}{123--130}.
\newblock


\bibitem[\protect\citeauthoryear{Sun and Zhang}{Sun and Zhang}{2018}]%
        {Sun:2018:CRS:3209978.3210002}
\bibfield{author}{\bibinfo{person}{Yueming Sun} {and} \bibinfo{person}{Yi
  Zhang}.} \bibinfo{year}{2018}\natexlab{}.
\newblock \showarticletitle{Conversational Recommender System}. In
  \bibinfo{booktitle}{\emph{SIGIR}}. \bibinfo{pages}{235--244}.
\newblock


\bibitem[\protect\citeauthoryear{Sutton, McAllester, Singh, and Mansour}{Sutton
  et~al\mbox{.}}{2000}]%
        {REINFORCE}
\bibfield{author}{\bibinfo{person}{Richard~S Sutton}, \bibinfo{person}{David~A
  McAllester}, \bibinfo{person}{Satinder~P Singh}, {and}
  \bibinfo{person}{Yishay Mansour}.} \bibinfo{year}{2000}\natexlab{}.
\newblock \showarticletitle{Policy gradient methods for reinforcement learning
  with function approximation}. In \bibinfo{booktitle}{\emph{NeurIPS}}.
  \bibinfo{pages}{1057--1063}.
\newblock


\bibitem[\protect\citeauthoryear{Wang, Wu, and Wang}{Wang
  et~al\mbox{.}}{2017}]%
        {wang2017factorization}
\bibfield{author}{\bibinfo{person}{Huazheng Wang}, \bibinfo{person}{Qingyun
  Wu}, {and} \bibinfo{person}{Hongning Wang}.} \bibinfo{year}{2017}\natexlab{}.
\newblock \showarticletitle{Factorization Bandits for Interactive
  Recommendation.}. In \bibinfo{booktitle}{\emph{AAAI}}.
  \bibinfo{pages}{2695--2702}.
\newblock


\bibitem[\protect\citeauthoryear{Wang, Huang, Xu, Shen, and Nie}{Wang
  et~al\mbox{.}}{2018}]%
        {sigir18/chatmore}
\bibfield{author}{\bibinfo{person}{Wenjie Wang}, \bibinfo{person}{Minlie
  Huang}, \bibinfo{person}{Xin-Shun Xu}, \bibinfo{person}{Fumin Shen}, {and}
  \bibinfo{person}{Liqiang Nie}.} \bibinfo{year}{2018}\natexlab{}.
\newblock \showarticletitle{Chat More: Deepening and Widening the Chatting
  Topic via A Deep Model}. In \bibinfo{booktitle}{\emph{SIGIR}}.
  \bibinfo{pages}{255--264}.
\newblock


\bibitem[\protect\citeauthoryear{Wu, Iyer, and Wang}{Wu et~al\mbox{.}}{2018}]%
        {wu2018learning}
\bibfield{author}{\bibinfo{person}{Qingyun Wu}, \bibinfo{person}{Naveen Iyer},
  {and} \bibinfo{person}{Hongning Wang}.} \bibinfo{year}{2018}\natexlab{}.
\newblock \showarticletitle{Learning Contextual Bandits in a Non-stationary
  Environment}. In \bibinfo{booktitle}{\emph{SIGIR}}.
  \bibinfo{pages}{495--504}.
\newblock


\bibitem[\protect\citeauthoryear{Wu, Wang, Gu, and Wang}{Wu
  et~al\mbox{.}}{2016}]%
        {wu2016contextual}
\bibfield{author}{\bibinfo{person}{Qingyun Wu}, \bibinfo{person}{Huazheng
  Wang}, \bibinfo{person}{Quanquan Gu}, {and} \bibinfo{person}{Hongning Wang}.}
  \bibinfo{year}{2016}\natexlab{}.
\newblock \showarticletitle{Contextual bandits in a collaborative environment}.
  In \bibinfo{booktitle}{\emph{SIGIR}}. ACM, \bibinfo{pages}{529--538}.
\newblock


\bibitem[\protect\citeauthoryear{Yu, Shen, and Jin}{Yu et~al\mbox{.}}{2019}]%
        {yu2019visual}
\bibfield{author}{\bibinfo{person}{Tong Yu}, \bibinfo{person}{Yilin Shen},
  {and} \bibinfo{person}{Hongxia Jin}.} \bibinfo{year}{2019}\natexlab{}.
\newblock \showarticletitle{An Visual Dialog Augmented Interactive Recommender
  System}. In \bibinfo{booktitle}{\emph{SIGKDD}}. ACM,
  \bibinfo{pages}{157--165}.
\newblock


\bibitem[\protect\citeauthoryear{Zhang, Xie, Li, and Lui}{Zhang
  et~al\mbox{.}}{2019}]%
        {zhang2019toward}
\bibfield{author}{\bibinfo{person}{Xiaoying Zhang}, \bibinfo{person}{Hong Xie},
  \bibinfo{person}{Hang Li}, {and} \bibinfo{person}{John Lui}.}
  \bibinfo{year}{2019}\natexlab{}.
\newblock \showarticletitle{Toward Building Conversational Recommender Systems:
  A Contextual Bandit Approach}.
\newblock \bibinfo{journal}{\emph{arXiv preprint arXiv:1906.01219}}
  (\bibinfo{year}{2019}).
\newblock


\bibitem[\protect\citeauthoryear{Zhang, Chen, Ai, Yang, and Croft}{Zhang
  et~al\mbox{.}}{2018}]%
        {zhang2018towards}
\bibfield{author}{\bibinfo{person}{Yongfeng Zhang}, \bibinfo{person}{Xu Chen},
  \bibinfo{person}{Qingyao Ai}, \bibinfo{person}{Liu Yang}, {and}
  \bibinfo{person}{W~Bruce Croft}.} \bibinfo{year}{2018}\natexlab{}.
\newblock \showarticletitle{Towards conversational search and recommendation:
  System ask, user respond}. In \bibinfo{booktitle}{\emph{CIKM}}.
  \bibinfo{pages}{177--186}.
\newblock


\end{thebibliography}

\end{document}